\documentstyle [11pt,epsf] {article}

\catcode`@=12
\newcommand{\be}{\begin{equation}}
\newcommand{\ee}{\end{equation}}
\newcommand{\ber}{\begin{eqnarray}}
\newcommand{\eer}{\end{eqnarray}}
\newcommand{\bers}{\begin{eqnarray*}}
\newcommand{\eers}{\end{eqnarray*}}
\newcommand{\tpi}{{\tilde{\pi}}}
\begin{document}
\vspace{0.5in}
\oddsidemargin -.375in  
\newcount\sectionnumber 
\sectionnumber=0 

\def\be{\begin{equation}} 
\def\ee{\end{equation}}
\thispagestyle{empty}
\begin{flushright} UH-511-921-98 \\
UT-PT-98-17\\ December 1998\\
\end{flushright}
\vspace {.5in}
\begin{center}
{\Large \bf{Flavor Changing Processes in Quarkonium Decays\\}}
\vspace{.5in}
{\rm {A. Datta$^3$,  
 P. J. O'Donnell$^3$, S. Pakvasa$^2$   
and X. Zhang$^{1, 4}$} \\}

\vskip .3in
{\it $^1$ CCAST (World Laboratory), P.O. Box 8730, Beijing 100080 and \\}
{ \it Institute of High Energy Physics\\}
{\it Academia Sinica, Beijing 100039, China\\}

{\it $^2$Department of Physics and Astronomy \\}
{\it  University of Hawaii, \\}
{\it Honolulu, Hawaii 96822\\}

{\it $^3$Department of Physics  \\}
{\it  University of Toronto, \\}
{\it Toronto, Ontario, Canada M5S 1A7\\}
 
{\it and\\}
{\it $^4$ High Energy Section \\}
{\it ICTP, 34014 Trieste, Italy\\}

%\vspace{.1in}
\vskip .5in
\end{center}  
\begin{abstract}
We study flavor changing processes
$\Upsilon \rightarrow B/{\overline B} X_s$ and 
$J/\psi \rightarrow D/{\overline D} X_u$ in the B factories and 
the Tau-Charm factories. In the standard model, these processes are predicted to be 
unobservable, so they serve as a probe of the new physics.
We first perform a model independent analysis, then examine the predictions of 
models; such as TopColor models, MSSM with R-parity
violation  and the two Higgs doublet model; for the branching ratios of
$\Upsilon \rightarrow B/{\overline B} X_s$ and $J/\psi \rightarrow D/{\overline D} X_u$ 
. We find that these branching ratios 
 could be as large as $10^{-6}$ and $10^{-5}$ in the presence of new physics.
 \end {abstract}
\newpage
\baselineskip 24pt

%
% Turn this off to get double spaced version
%\tighten
%
%

\section{Introduction}
The possibility of observing large CP violating asymmetries in the
decay of $B$ mesons has motivated the construction of high luminosity $B$
factories at several of the world's high energy physics laboratories. 
These $B$ factories will be producing roughly about $10^8$ Upsilons.
Meanwhile BES has already accumulated $9 \times 10^6$ $J/\psi$
and plans to increase the number to $5\times 10^7$ in the near future.
An interesting question, that we  investigate in this
paper, is whether the large sample of the $\Upsilon$ and the $J/\psi$ can be used to
probe flavor changing processes in the decays of $\Upsilon$ and $J/\psi$.
 In particular we  look at the
flavor changing processes $\Upsilon \to B/{\overline B} X_s$
and $J /\psi \to D/{\overline D} X_u$,   from the
 underlying 
$b \to s$ and $c \to u $ quark transitions. For the quarkonium 
system, these flavor changing processes 
are expected to be much smaller than in the case of decays of
 the $B$ or the $D$
meson
because of the larger decay widths of the bottonium and the charmonium systems
which decay via the strong interactions. Indeed the standard model contributions
to $\Upsilon \to B/{\overline B} X_s$
and $J /\psi \to D/{\overline D} X_u$ are tiny. However, new physics may
enhance the branching ratios for these processes. Whether 
this enhancement maybe sufficient for these
processes to be observable in the next round of experiments is the subject of this
work. Invisible decays of $\Upsilon$ and $J/\psi$ resonances in the standard model and beyond have been studied recently\cite{Ng}.

 Non leptonic decays of
heavy quarkonium systems can be more reliably calculated
 than the non leptonic decays of the heavy mesons. A consistent 
and systematic
formalism to handle heavy quarkonium decays is 
available in NRQCD \cite{lepage} which
is missing for the heavy mesons. As in the meson system 
\cite{datta1}        
it is more fruitful to concentrate 
on quasi-inclusive  processes like
$\Upsilon \to B/{\overline B} X_s$
and $J /\psi \to D/{\overline D} X_u$ because they 
can be calculated with  less theoretical uncertainty and have larger branching ratios
 than the purely
exclusive quarkonium  non leptonic decays. The branching ratios of
exclusive flavor changing non leptonic decays
of $\Upsilon$ and $J/\psi$ in the standard model
 have been calculated and found to be  very small \cite{indianguys}.

We begin with a model independent description of the processes
$\Upsilon \to B/{\overline B} X_s$
and $J /\psi \to D/{\overline D} X_u$. In the standard model
these decays can proceed through tree and penguin processes.
 For new
physics contribution to these processes 
we concentrate on four quark operators
of the type ${\overline s}b{\overline b} b$ and  
${\overline u}c{\overline c} c $. 
We choose the currents in the four quark operators to be scalars and so
these operators may arise through the exchange of a heavy scalar 
for e.g a Higgs or 
a leptoquark in some model of new physics. These four quark operators, 
at the one loop level, generate effective
${\overline s}b \{g,\gamma,Z \}$ and
${\overline u}c \{g,\gamma,Z \}$ vertices
 which 
would effect the flavor changing decays of
the $B$ and the $D$ mesons. The effective vertices for an on shell $g$ and $\gamma$
vanish and so there is no contribution to $b \to s \gamma$ or $c \to u \gamma$. We can 
however put constraints on these operators by considering the processes
$b \to s l^+ l^{-}$ and $c \to u l^+ l^{-}$.  
The constrained operators can
then be used to 
calculate the branching ratios for $\Upsilon \to B/ {\overline B} X_s$
and $J /\psi \to D/ {\overline D} X_u$. 

We then consider some models that may generate the kind of 
four quark operators described above. A few examples of models
where these operators can be generated are
top color models,  MSSM with R parity violation and  a general two 
Higgs doublet model without any discrete symmetry. In some cases constraints on
the parameters that appear in the prediction for the branching ratios for
$\Upsilon \to B/{\overline B} X_s$
and $J /\psi \to D/{\overline D} X_u$ are already available. In other cases the parameters 
are constrained, as in our  model independent analysis, from the processes
$ b \to s l^+ l^{-}$ and $c \to u l^+ l^{-}$. 

In the sections which follow, we describe the effective Hamiltonian
for the $\Upsilon \to B/{\overline B} X_s$
and $J /\psi \to D/{\overline D} X_u$. Next we describe the calculation of the
matrix elements and decay rates for these processes. We then discuss the
calculation of the effective
${\overline s}b \{g,\gamma,Z \}$ and
${\overline u}c \{g,\gamma,Z \}$ vertices and constraints from the processes
$ b \to s l^+ l^{-}$ and $c \to u l^+ l^{-}$. 
This is followed by a description of some models that
can generate the new four quark operators in the effective Hamiltonian
for $\Upsilon \to B/ {\overline B} X_s$
and $J /\psi \to D/ {\overline D} X_u$. Finally we present our
results and conclusions.

\section{ Effective Hamiltonian} 
In this section we  present the effective Hamiltonian for $\Upsilon$ 
decays. The effective Hamiltonian for charmonium decays can be  
written down by making obvious changes.
In the Standard Model (SM) 
the amplitudes for hadronic $\Upsilon$ decays of the type $b\bar{b}\to s
\bar{b} + \bar{s}b$ 
are generated by the following effective 
Hamiltonian \cite{Reina,buras}:
\begin{eqnarray}
H_{eff}^q &=& {G_F \over \protect \sqrt{2}} 
   [V_{fb}V^*_{fq}(c_1O_{1f}^q + c_2 O_{2f}^q) -
     \sum_{i=3}^{10}(V_{ub}V^*_{uq} c_i^u
+V_{cb}V^*_{cq} c_i^c +V_{tb}V^*_{tq} c_i^t) O_i^q] +H.C.\;,
\end{eqnarray}
where the
superscript $u,\;c,\;t$ indicates the internal quark, $f$ can be $u$ or 
$c$ quark, $q$ can be either a $d$ or a $s$ quark depending on 
whether the decay is a $\Delta S = 0$
or $\Delta S = -1$ process.
The operators $O_i^q$ are defined as
\begin{eqnarray}
O_{1f}^q &=& \bar q_\alpha \gamma_\mu Lf_\beta\bar
f_\beta\gamma^\mu Lb_\alpha\;,\;\;\;\;\;\;O_{2f}^q =\bar q
\gamma_\mu L f\bar
f\gamma^\mu L b\;,\nonumber\\
O_{3,5}^q &=&\bar q \gamma_\mu L b
\bar q' \gamma_\mu L(R) q'\;,\;\;\;\;\;\;\;O_{4,6}^q = \bar q_\alpha
\gamma_\mu Lb_\beta
\bar q'_\beta \gamma_\mu L(R) q'_\alpha\;,\\
O_{7,9}^q &=& {3\over 2}\bar q \gamma_\mu L b  e_{q'}\bar q'
\gamma^\mu R(L)q'\;,\;O_{8,10}^q = {3\over 2}\bar q_\alpha
\gamma_\mu L b_\beta
e_{q'}\bar q'_\beta \gamma_\mu R(L) q'_\alpha\;,\nonumber
\end{eqnarray}
where $R(L) = 1 \pm \gamma_5$, 
and $q'$ is summed over all flavors except t.  $O_{1f,2f}$ are the tree
level and QCD corrected operators. $O_{3-6}$ are the strong gluon induced
penguin operators, and operators 
$O_{7-10}$ are due to $\gamma$ and Z exchange (electroweak penguins),
and ``box'' diagrams at loop level. The Wilson coefficients
 $c_i^f$ are defined at the scale $\mu \approx m_b$ 
and have been evaluated to next-to-leading order in QCD.
The $c^t_i$ are the regularization scheme 
independent values obtained in Ref. \cite{FSHe}.
We give the non-zero  $c_i^f$ 
below for $m_t = 176$ GeV, $\alpha_s(m_Z) = 0.117$,
and $\mu = m_b = 5$ GeV,
\begin{eqnarray}
c_1 &=& -0.307\;,\;\; c_2 = 1.147\;,\;\;
c^t_3 =0.017\;,\;\; c^t_4 =-0.037\;,\;\;
c^t_5 =0.010\;,
 c^t_6 =-0.045\;,\nonumber\\
c^t_7 &=&-1.24\times 10^{-5}\;,\;\; c_8^t = 3.77\times 10^{-4}\;,\;\;
c_9^t =-0.010\;,\;\; c_{10}^t =2.06\times 10^{-3}\;, \nonumber\\
c_{3,5}^{u,c} &=& -c_{4,6}^{u,c}/N_c = P^{u,c}_s/N_c\;,\;\;
c_{7,9}^{u,c} = P^{u,c}_e\;,\;\; c_{8,10}^{u,c} = 0
\end{eqnarray}
where $N_c$ is the number of color. 
The leading contributions to $P^i_{s,e}$ are given by:
 $P^i_s = ({\frac{\alpha_s}{8\pi}}) c_2 ({\frac{10}{9}} +G(m_i,\mu,q^2))$ and
$P^i_e = ({\frac{\alpha_{em}}{9\pi}})
(N_c c_1+ c_2) ({\frac{10}{9}} + G(m_i,\mu,q^2))$.  
The function
$G(m,\mu,q^2)$ is given by
\begin{eqnarray}
G(m,\mu,q^2) = 4\int^1_0 x(1-x)  \mbox{ln}{m^2-x(1-x)q^2\over
\mu^2} ~\mbox{d}x \;.
\end{eqnarray}
All the above coefficients are obtained up to one loop order in electroweak 
interactions. The momentum $q$ is the momentum carried by the virtual gluon in
the penguin diagram.
When $q^2 > 4m^2$, $G(m,\mu,q^2)$ becomes imaginary. 
In our calculation, we 
use $m_u = 5$ MeV, $m_d = 7$ MeV, $m_s = 200$ MeV, $m_c = 1.35$ GeV
\cite{lg,PRD}.
For $\Upsilon \to {\overline B} X_s$, the operators $O_{1f}^q$ and
$O_{2f}^q$ do not contribute and the gluon momentum
is fixed at  $q^2=M^2_{\Upsilon}$.

A similar expression for the standard model contribution to
the flavor changing decays  $J/\psi$ can be written down.

To the standard model contribution we add higher dimensional 
four quark operators generated by physics beyond the standard model.
In this paper, we consider the four quark operators with two scalar
currents.
\ber
L_{new} & = &\frac{R_1}{\Lambda^2}{\overline s}(1-\gamma^5)b{\overline b} 
(1+\gamma^5) b +
\frac{R_2}{\Lambda^2} {\overline s}(1+\gamma^5)b{\overline b}
(1-\gamma^5) b +h.c . \
\eer
\noindent
%The four quark operators in $L_{new}$ are the product of two scalar currents.
The four quark operators in $L_{new}$ are the product of two scalar currents.
In Eq. (5) $\Lambda$ represents the new physics scale and $R_1$ and $R_2$ are two free
parameters which describe the strength of the contribution of the underlying new physics
to the effective operators. In our analysis we will only keep dimension 
six operators suppressed by $1/\Lambda^2$ and neglect all higher 
dimension operators.

Using a Fierz transformation one can express the scalar-scalar combination
in terms of vector-vector combination. For instance we can write
\ber
{\overline s}_{\alpha}(1-\gamma^5)b_{\alpha}
{\overline b}_{\beta}(1+\gamma^5)b_{\beta} & = &
-\frac{1}{2}{\overline s}_{\alpha}\gamma_{\mu}(1+\gamma^5)b_{\beta}
{\overline b}_{\beta}\gamma^{\mu}(1-\gamma^5)b_{\alpha} \nonumber\\
& =&
-\frac{1}{2N_c}{\overline s}_{\alpha}\gamma_{\mu}(1+\gamma^5)b_{\alpha}
{\overline b}_{\beta}\gamma^{\mu}(1-\gamma^5)b_{\beta} \nonumber\\
& - &{\overline s}_{\alpha}T^a_{\alpha \beta}\gamma_{\mu}(1-\gamma^5)b_{\beta}
{\overline b}_{\beta^{\prime}}
T^a_{\beta^{\prime} \alpha^{\prime}}\gamma^{\mu}(1-\gamma^5)b_{\alpha^{\prime}}
\
\eer
where $T^a$ are the $SU(3)$ color matrices with the normalization
$ Tr[T^aT^b]=\delta_{ab}/2 $ and $N_c$ is the number of colors.
In the quarkonium system the leading component in the Fock space expansion involves
the quark-antiquark pair being in a $^3S_1$ state,  probed
by the operator
${\overline s}\gamma_{\mu}(1+\gamma^5)b
{\overline b}\gamma^{\mu}(1-\gamma^5)b$.
Note the general Fock space expansion of  quarkonium
in NRQCD is \cite{sean}
\ber
| \psi_Q > &= & O(1) |\, {\overline Q}Q\,[{}^3S_1^{(1)}] \rangle
+ O(v) |\, {\overline Q}Q\, [ {}^3P_J^{(8)}] \,  g \rangle \nonumber\\
&  + & O(v^2) |\, {\overline Q}Q\, [ {}^3S_1^{(1,8)}] \,  g g \rangle
+  O(v^2) |\, {\overline Q}Q\, [ {}^1S_0^{(8)}] \,  g \rangle
+ O(v^2) |\, {\overline Q}Q\, [ {}^3D_J^{(1,8)}] \,  g g \rangle + \cdots 
, \
\eer
where $v$ is the  velocity of the constituents in the quarkonium and
 $g$ represents a dynamical gluon, {\it i.e.} one whose effects cannot be
incorporated into an instantaneous potential and whose typical momentum is
$m_Q v^2$.   
The low energy hadronization of the leading component in the Fock space
expansion of the quarkonium takes place at $O(v^3)$. 
 As to the other Fock states notice that the
$|\, {\overline Q}Q\, [ {}^3P_J^{(8)}] \,  g \rangle$,
 configuration arises when the predominant state radiates a soft dynamical
gluon. Such a process is mediated principally by the electric dipole operator,
for which the selection rule is $L' = L\pm 1$, $S' = S$, and which involves
a single power of heavy quark three-momentum. Thus, the coefficient
associated with this state is of order $v$.  
The electric dipole emission of yet
another gluon involves a change from the $P$-wave state
to the $S$- and $D$-wave states
$|\, {\overline Q}Q\, [ {}^3S_1^{(1,8)}] \,  g g \rangle $,
$|\, {\overline Q}Q\rangle, [ {}^3D_J^{(1,8)}] \,  g g \rangle $
and so the associated  coefficients 
 are of order $v^2$.  Finally,  the coefficient
of the state $|\, {\overline Q}Q\, [ {}^1S_0^{(8)}] \,  g \rangle $ results
from fluctuations
into this spin-singlet state from the predominant spin-triplet state
with the emission of a soft gluon via a spin-flipping
magnetic dipole transition. Such transitions involve
the gluon three-momentum
$(\sim m_Q v^2)$
rather than the heavy quark three-momentum
$(\sim m_Q v)$,
and therefore the associated coefficient  is of order $v^2$.
The low energy hadronization of these component in the Fock space
expansion of the quarkonium takes place at $O(v^7)$. Also  for the
 P wave, an additional
factor of $v$ comes from the derivative of the wavefunction.

Before concluding this sections, we point out that 
besides operator in Eq. (5) there are
additional four quark operators in the effective 
lagrangian\cite{wyler}, such as
those with vector-vector current structure, which
contribute also to the processes we consider. We focus on operators in Eq. (5)
because, as mentioned earlier, they 
can be generated by the
exchanges of the
new scalar bosons in models we consider below in section 5.

\section{Matrix Elements for $ \Upsilon  \rightarrow {\overline B}  X_s $}

We proceed to calculate 
 the matrix elements of the form 
$<{\overline B} X_s|H_{eff}|\Upsilon>$ which represents 
the process  $ \Upsilon\rightarrow {\overline B} X_s$ and where $H_{eff}$ has been 
described 
above. The effective Hamiltonian consists of operators with a
 current $\times$ current structure. 
Pairs of such operators can be expressed in terms of
color singlet and color octet structures. The factorization formalism based on
 NRQCD \cite{lepage}, which allows a systematic and consistent 
probe of the complete quarkonium Fock space,
 can then be used to calculate the $\Upsilon$ decay rate.

The matrix element of $\Upsilon \rightarrow {\overline B} X_s$ decay,   
can be expressed as, 

\ber
	M & = &\frac{G_F}{\protect \sqrt{2}}W_1
 <{\overline B} X_s|\, \overline{s} \gamma^\mu(1-\gamma^5) \, b\, |\, 0>
	      <0|{\overline {b}}\gamma_{\mu}b|\Upsilon>\nonumber  \\
          &+&\frac{G_F}{\protect \sqrt{2}}W_1^{\prime}
 <{\overline B} X_s|\, \overline{s} \gamma^\mu(1+\gamma^5) \, b\, |\, 0>
	      <0|{\overline {b}}\gamma_{\mu}b|\Upsilon>\nonumber  \\
	 & + &\frac{G_F}{\protect \sqrt{2}} W_8
         <{\overline B} X_s|\, \overline{s} 
         \gamma^\mu(1-\gamma^5)T^a \, b\, |\, 0>
	      <0|{\overline {b}}\gamma_{\mu}T^ab|\Upsilon>\nonumber  \\
         & + &\frac{G_F}{\protect \sqrt{2}} W_8^{\prime}
         <{\overline B} X_s|\, \overline{s} 
         \gamma^\mu(1+\gamma^5)T^a \, b\, |\, 0>
	      <0|{\overline {b}}\gamma_{\mu}T^ab|\Upsilon>\nonumber  \\
        & + &\frac{G_F}{\protect \sqrt{2}} U_8
         <{\overline B} X_s|\, \overline{s} 
         \gamma^\mu(1-\gamma^5)T^a \, b\, |\, 0>
	      <0|{\overline {b}}\gamma_{\mu}\gamma^5T^ab|\Upsilon>\nonumber  \\
         & + &\frac{G_F}{\protect \sqrt{2}} U_8^{\prime}
         <{\overline B} X_s|\, \overline{s} 
         \gamma^\mu(1+\gamma^5)T^a \, b\, |\, 0>
	      <0|{\overline {b}}\gamma_{\mu}\gamma^5T^ab|\Upsilon>\nonumber  \\
              & + &\frac{G_F}{\protect \sqrt{2}} V_8
         <{\overline B} X_s|\, \overline{s} (1+\gamma^5)T^a \, b\, |\, 0>
	      <0|{\overline {b}}(1-\gamma^5)T^ab|\Upsilon> \nonumber\\
         & + &\frac{G_F}{\protect \sqrt{2}} V_8^{\prime}
         <{\overline B} X_s|\, \overline{s} (1-\gamma^5)T^a \, b\, |\, 0>
	      <0|{\overline {b}}(1+\gamma^5)T^ab|\Upsilon> \
\eer
where
\ber
  W_1& = &W_{1std} +W_{1new} \nonumber  \\
  W_1^{\prime} &=&W_{1std}^{\prime} +W_{1new}^{\prime}\nonumber  \\
  W_8&=&W_{8std} +W_{8new} \nonumber  \\
  W_8^{\prime}&=& W_{8std}^{\prime} +W_{8new}^{\prime}\nonumber  \\
  U_8 & = & U_{8std} +U_{8new}\nonumber  \\
  U_8^{\prime} & = & U_{8std}^{\prime} +U_{8new}^{\prime}\nonumber  \\
  V_8 & = & V_{8std} +V_{8new}\nonumber\\
  V_8^{\prime} & = & V_{8std}^{\prime} +V_{8new}^{\prime}\
\eer
with
\ber
        W_{1std} & = &  \left[\{A_3 +A_4 -\frac{1}{2}(A_9 +A_{10})\}(1+ 
		 \frac{1}{N_c}) +(A_5+\frac{A_6}{N_c}) 
                  -\frac{1}{2}(A_7+\frac{A_8}{N_c})\right]  \nonumber  \\
        W_{1std}^{\prime}&=& 0 \nonumber  \\ 
	W_{8std} & = & \left[2(A_3 +A_4 +A_6)-(A_8+A_9+A_{10})\right]
        \nonumber \\
        W_{8std}^{\prime}& = & 0 \nonumber  \\ 
	U_{8std} & = &\left[2(-A_3 -A_4 +A_6)-(A_8-A_9-A_{10})\right] 
        \nonumber \\
        U_{8std}^{\prime} & = &0 \nonumber\\
	V_{8std} & = & \left[-4A_5 +2A_7\right] \nonumber\\
        V_{8std}^{\prime} & = & 0 \
\eer

 and
\ber
	W_{1new} & = & -\frac{1}{2N_c}\frac{\sqrt{2}}{G_F}
        \frac{R_2}{\Lambda^2} \nonumber  \\ 
	W_{1new}^{\prime} & = &-\frac{1}{2N_c}\frac{\sqrt{2}}{G_F}
        \frac{R_1}{\Lambda^2} \nonumber  \\ 
        W_{8new} & = & -\frac{\sqrt{2}}{G_F}
        \frac{R_2}{\Lambda^2} \nonumber  \\ 
	W_{8new}^{\prime} & = &-\frac{\sqrt{2}}{G_F}
        \frac{R_1}{\Lambda^2} \nonumber  \\  
	U_{8new} & = & -\frac{\sqrt{2}}{G_F}
        \frac{R_2}{\Lambda^2} \nonumber  \\ 
        U_{8new}^{\prime} & = &\frac{\sqrt{2}}{G_F}
        \frac{R_1}{\Lambda^2} \nonumber  \\  
        V_{8}^{\prime} &= & 0 \
\eer
We have defined
\be
	A_i = - \sum_{q=u,c,t} c_i^q V_q
\ee
with
\be
	V_q = V_{qs}^{*} V_{qb}
\ee

Similar expressions can be written for the matrix elements describing
the $J/\psi$ decay.

To calculate the decay rate we use the parton model  to write the process
 $  \Upsilon \to {\overline B} X_s $  as
  $  \Upsilon(P) \to {\overline b(p_1)} s(p_2) $ 
The squared matrix element  is then given by
\ber
|M|^2 & = & 2MZ_1\left[ <\Upsilon|O_1({}^3S_1)|\Upsilon>(|W_1|^2 +
|{W_1^{\prime}}^2|)+
(<\Upsilon|O_8({}^3S_1)+ 2\frac{O_8({}^3P_1)}{m_b^2}
|\Upsilon>)(|W_8|^2 +
|{W_8^{\prime}}^2|)\right]\nonumber\\
& + & 6MZ_2\left[ 
(<\Upsilon|O_8({}^1S_0)|\Upsilon>(|U_8|^2 +
|{U_8^{\prime}}^2|)\right]\nonumber\\
& + & 6MZ_3\left[ 
(<\Upsilon|O_8({}^1S_0)|\Upsilon>(|V_8|^2 +
|{V_8^{\prime}}^2|)\right]\
\eer
where
\ber
Z_1 &=& 8\left[p_1 \cdot p_2 +\frac{2p_1 \cdot P p_2 \cdot P}{M^2}\right]
\nonumber\\
Z_2 &=& 8\left[-p_1 \cdot p_2 +\frac{2p_1 \cdot P p_2 \cdot P}{M^2}\right]
\nonumber\\
Z_3 & = & 8\left[p_1 \cdot p_2\right]\
\eer
with $M$ being the quarkonium mass.
The matrix elements of the various color singlet and color octet operators, 
$O_1(^{(2S+1)}L_J)$ and
$O_8(^{(2S+1)}L_J)$ encode the non perturbative long distance effects in the
evolution of ${\overline Q}{Q}(^{(2S+1)}L_J)_{1,8}$ to $\Upsilon$.

Along with the CP violating phases present in the standard model contribution
there can be additional phases from the new contribution. We can then
 construct the CP violating rate asymmetry as
\ber
a_{CP}=\frac{\Gamma(\Upsilon \to {\overline B}X_s)-
\Gamma(\Upsilon \to  B {\overline X_s})}
{\Gamma(\Upsilon \to {\overline B}X_s)+
\Gamma(\Upsilon \to  B {\overline X_s})}\
\eer
\section{Low Energy Constraints}
The lagrangian $L_{new}$ generates, at one loop level, the effective
${\overline s}b\gamma ^*$,
${\overline s}bg^*$,${\overline s}b Z$ vertices as shown in Fig. 1, where
 $\gamma^*$ and $g^*$ indicate an off shell photon and a gluon..
Similar vertices involving $c \to u$ transitions 
 are generated in the charmonium sector also. These 
vertices, with a $\gamma$ and $Z$, will contribute to $b \to s l^+l^-$ and
$c \to u l^+l^-$. Note there is no contribution to $b \to s \gamma$. The vertex
$b \to s g^*$ can give rise to the process $b \to s q {\overline q}$ which 
will contribute to non-leptonic $B$ decays. We expect the constraints from
$b \to s l^+l^-$ to be better than from non-leptonic $B$ decays 
because of the theoretical uncertainties in calculating non-leptonic decays.
\begin{figure}[htb]
\vskip 0.9in
\centerline{\epsfysize 3.0 truein \epsfbox{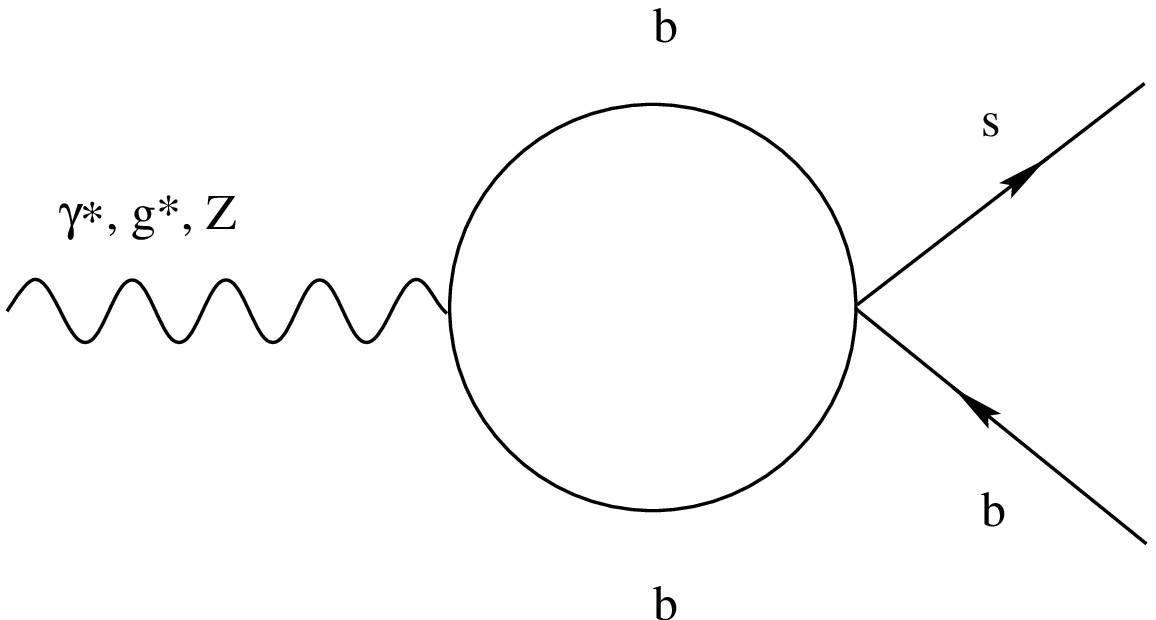}}
\caption{Effective ${\overline s}b\gamma^*$,
${\overline s}bg^*$,${\overline s}b Z$ vertices generated by $L_{new}$
}
\label{bkx_rate}
\end{figure}
The effective ${\overline s}b\gamma^*$,
${\overline s}bg^*$,${\overline s}b Z$ lagrangian can be written as
\ber
\delta L_{{\overline s} b \gamma^*} & = &\frac{e_b}{\Lambda^2}
{\overline s}
\left[R_{+}C_{+}^{\mu} + R_{-}C_{-}^{\mu}\right]b\
\eer
where
\ber
R_{+} &= &\frac{R_1 + R_2}{2}\nonumber\\
R_{-}& = &\frac{R_1 - R_2}{2}\nonumber\\
C_{+} & = & \frac{1}{16\pi^2}\int^1_0 dx \log\left(\frac{\Lambda^2}{B^2}\right)
\left[8q^{\mu}\gamma \cdot q x(x-1) + 8\gamma^{\mu}q^2 x(1-x)\right] 
\nonumber\\
C_{-} & = & \frac{1}{16\pi^2}\int^1_0 dx \log\left(\frac{\Lambda^2}{B^2}\right)
\left[8q^{\mu}\gamma \cdot q \gamma^5x(x-1) + 
8\gamma^{\mu}\gamma^5q^2 x(1-x)\right]\ 
\eer
with 
\bers
B^2 &= & m_b^2 -q^2x(1-x)\
\eers
where $e_b$ is the $b$ quark electric charge and $q$ is the photon momentum.
The effective ${\overline s}b g^*$ lagrangian can be obtained by 
replacing $e_b$ by $g_s$, the strong coupling constant.
The effective ${\overline s}b Z$ lagrangian can be written as
\ber
\delta L_{{\overline s} b Z} & = &\frac{g}{2c_w\Lambda^2}
 {\overline s}\left[R_{+}(g_LF_{1+} +g_RF_{2+})+
R_{-}(g_LF_{1-} +g_RF_{2-})\right]\
\eer
where
\ber
F_{1+} & = & \frac{1}{16\pi^2}\int^1_0 dx \log \left(\frac{\Lambda^2}{B^2}\right)
\left[8q^{\mu}\gamma \cdot q x(x-1)(1+\gamma^5)+ 
8\gamma^{\mu}q^2 x(1-x)(1+\gamma^5) -8\gamma^{\mu}\gamma^5m_b^2\right]
\nonumber\\ 
F_{1-} & = & \frac{1}{16\pi^2}\int^1_0 dx 
\log \left(\frac{\Lambda^2}{B^2}\right)
\left[8q^{\mu}\gamma \cdot q x(x-1)(1+\gamma^5) 
8\gamma^{\mu}q^2 x(1-x)(1+\gamma^5) -8\gamma^{\mu}m_b^2\right]
\nonumber\\ 
F_{2+} & = & \frac{1}{16\pi^2}\int^1_0 dx \log \left(\frac{\Lambda^2}{B^2}\right)
\left[8q^{\mu}\gamma \cdot q x(x-1)(1-\gamma^5) 
8\gamma^{\mu}q^2 x(1-x)(1-\gamma^5) +8\gamma^{\mu}\gamma^5m_b^2\right]
\nonumber\\ 
F_{2-} & = & \frac{1}{16\pi^2}\int^1_0 dx \log \left(\frac{\Lambda^2}{B^2}\right)
\left[8q^{\mu}\gamma \cdot q x(1-x)(1-\gamma^5) 
8\gamma^{\mu}q^2 x(x-1)(1-\gamma^5) +8\gamma^{\mu}m_b^2\right] \
\eer
with
\bers
g_L &=&-\frac{1}{2} +\frac{1}{3}s^2_w \nonumber\\ 
g_R & = &\frac{1}{3}s^2_w \
\eers

For $b \to s l^+ l^-$ 
the $q^{\mu}$ terms in the equations above do not 
contribute if we neglect the lepton masses. Furthermore 
the contribution from the $Z$ exchange is suppressed with respect 
to the $\gamma$ exchange by factor of $q^2/M_Z^2$ and so we do not include 
the $Z$ contribution. 
The additional contribution
to the effective Hamiltonian for  $b \to s l^+ l^-$ can be written as
\ber
\delta H_{b \to s l^+ l^-} & = & -\frac{e^2}{16\pi^2}\frac{e_b}{\Lambda^2}
\int^1_0 dx 8x(1-x)\log \left(\frac{\Lambda^2}{B^2}\right)
\left[R_1{\overline s}\gamma^{\mu}b_L{\overline l}\gamma_{\mu}l+   
R_2{\overline s}\gamma^{\mu}b_R{\overline l}\gamma_{\mu}l\right]\
\eer   
which has to be added to the standard model contribution \cite{buras}.
Similar results can also be written for the charm sector.
\section{Models}
In this section we look at various models that can give rise to 
$L_{new}$ given in Eq. 5.
As a first example we consider a recent version of
 top color models\cite{hill}. In such models the 
top quark participates in a new strong interaction which is broken at 
some high energy 
scale $\Lambda$. The strong interaction, though not confining, leads to the 
formation of a top condensate $<{\overline t}_Lt_R>$ resulting in a large
 dynamical mass for the top quark. The scale $\Lambda$ is chosen to 
be of the order of a TeV to avoid naturalness problem which implies that the 
electroweak symmetry cannot be broken solely by the top condensate.
In the low energy sector of the theory, scalar bound states are formed that 
couple strongly to the $b$ quark \cite{kominis,burdman}
\ber
L_b & = & \frac{m_t}{f_{\tpi}\sqrt{2}}{\overline b}_L(H +iA^0)b_R +h.c \
\eer
where $f_{\tpi} \sim 50$ GeV is the top pion decay constant. On 
integrating out the Higgs fields $H$ and $A^0$ we have an effective 
four fermion operator
\ber
L_{eff} & = &\frac{m_t^2}{f_{\tpi}^2m_H^2}
{\overline b}_Lb_R 
{\overline b}_Rb_L  \
\eer
Since the $b$ quark in (22) is in 
the weak-eigenstate, $L_{eff}$ in (22) will induce
flavor changing neutral current (FCNC) four quark operators in Eq. (5) after 
diagonalizing the quark mass matrix\cite{burdman},
with coefficients,
\ber
R_1 & = & \frac{1}{4}\frac{m_t^2}{f_{\tpi}^2m_H^2}
|D_{Lbb}|^2D_{Rbb}D^*_{Rbs}\nonumber\\
R_2 &=& \frac{1}{4}\frac{m_t^2}{f_{\tpi}^2m_H^2}
|D_{Rbb}|^2D_{Lbb}D^*_{Lbs} \
\eer
where $D_L$ and $D_R$ are the mixing matrices in the left and the 
right handed down sector. In the charm sector similar interactions can arise 
due to the strong couplings of the top quark to top pions. 
%On integrating out 
%the top pions one can obtain an similar interaction to Eq. 5. 
The effective operators generated by integrating out the top-pions are similar to Eq. (5)
with replacement of $b$ by $c$ and $s$ by $u$.
In topcolor II
models \cite{burdman,hill2}, where there can be strong top-pion
couplings of the top with the charm quark, we have 
\ber
R_1 & = & \frac{1}{4}\frac{m_t^2}{f_{\tpi}^2m_{\tpi}^2}
|U_{Lcc}|^2U_{Rtc}U^*_{Rtu}\nonumber\\
R_2 & = & \frac{1}{4}\frac{m_t^2}{f_{\tpi}^2m_{\tpi}^2}
|U_{Rtc}|^2U_{Lcu}U^*_{Lcc}\
\eer

\noindent
In supersymmetric standard models without $R$ parity,
%, the $R$-parity of a field with
%spin $S$, 
%baryon-number $B$ and  lepton-number $L$ is defined to be
%\begin{equation}
%R=(-1)^{2S+3B+L}.
%\end{equation}
%$R$ is +1 for all the SM particles and -1 for all super particles.
%$R$-parity invariance is often imposed on the Lagrangian in order to
%maintain the separate conservation of baryon-number and lepton-number.
% R-parity conservation,
%the conservation is not dictated by any fundamental principle such as
%gauge invariance and there is no compelling theoretical motivation for it. 
the most general superpotential of 
the MSSM,
consistent with $SU(3)\times SU(2)\times U(1)$
 gauge symmetry and supersymmetry, can be written as
\begin{equation}
{\cal W}={\cal W}_R+{\cal W}_{\not \! R},
\end{equation}
where ${\cal W}_R$ is the $R$-parity conserving part while 
${\cal W}_{\not \! R}$ violates the $R$-parity. They are given by 
\begin{eqnarray}
{\cal W}_R&=&h_{ij}L_iH_2E_j^c+h_{ij}^{\prime}Q_iH_2D_j^c
             +h_{ij}^{\prime\prime}Q_iH_1U_j^c,\\ \label{RV}
{\cal W}_{\not \! R}&=&\lambda_{ijk}L_iL_jE_k^c
+\lambda_{ijk}^{\prime}L_iQ_jD_k^c
             +\lambda_{ijk}^{\prime\prime}U_i^cD_j^cD_k^c+\mu_iL_iH_2.
\end{eqnarray}
Here $L_i(Q_i)$ and $E_i(U_i,D_i)$ are the left-handed
lepton (quark) doublet and lepton (quark) singlet chiral superfields, with
$i,j,k$ being generation indices and $c$ denoting a charge conjugate field.
$H_{1,2}$ are the  chiral superfields
representing the two Higgs doublets. 
In the  $R$-parity violating superpotential above, the  
$\lambda$ and $\lambda^{\prime}$ couplings 
violate lepton-number conservation, while the
$\lambda^{\prime\prime}$ couplings violate baryon-number conservation.
$\lambda_{ijk}$ is antisymmetric in the first two
indices and $\lambda^{\prime\prime}_{ijk}$ is antisymmetric in
the last two indices.
While it is theoretically possible to have both baryon-number 
and lepton-number violating terms in the lagrangian, the non-observation
of proton decay imposes very stringent conditions on their simultaneous
presence \cite{proy}.
We, therefore, assume the existence of either  $L$-violating couplings or  
$B$-violating couplings, but not the coexistence of both. 
We calculate the effects of both types of couplings.

In terms of the four-component Dirac notation, the  lagrangian involving the
$\lambda^{\prime}$ and $\lambda^{\prime\prime}$ couplings  is given by
\begin{eqnarray}
{\cal L}_{\lambda^{\prime}}&=&-\lambda^{\prime}_{ijk}
\left [\tilde \nu^i_L\bar d^k_R d^j_L+\tilde d^j_L\bar d^k_R\nu^i_L
       +(\tilde d^k_R)^*(\bar \nu^i_L)^c d^j_L\right.\nonumber\\
& &\hspace{1cm} \left. -\tilde e^i_L\bar d^k_R u^j_L
       -\tilde u^j_L\bar d^k_R e^i_L
       -(\tilde d^k_R)^*(\bar e^i_L)^c u^j_L\right ]+h.c.,\\
{\cal L}_{\lambda^{\prime\prime}}&=&-\lambda^{\prime\prime}_{ijk}
\left [\tilde d^k_R(\bar u^i_L)^c d^j_L+\tilde d^j_R(\bar d^k_L)^c u^i_L
       +\tilde u^i_R(\bar d^j_L)^c d^k_L\right ]+h.c.
\end{eqnarray}
The terms proportional to $\lambda$ are not relevant to our present 
discussion and will not be considered here.
 The exchange of sneutrinos with the 
$\lambda^{\prime}$ coupling will generate $L_{new}$
for $\Upsilon \to {\overline B}X_s$ with
\ber
R_1 & = &\frac{1}{4}
 \Sigma_{i}\frac{\lambda^{\prime }_{i32} \lambda^{\prime *}_{i33}}
{m_{\tilde {\nu_i}}^2}\nonumber\\
R_2 &=&\frac{1}{4}
 \Sigma_{i}\frac{\lambda^{\prime *}_{i23} \lambda^{\prime}_{i33}}
{m_{\tilde {\nu_i}}^2}\
\eer
For the case of $J/\psi \rightarrow D X_u$ the operators in $L_{new}$ cannot be generated
at tree level.
 
Another model of interest is an extension of the
  SM with  additional scalar SU(2) doublets, the simplest of these would be
the two Higgs doublet model (2HDM). In general, when the quarks couple to more than
one scalar doublet, there are inevitably FCNC
couplings to the neutral scalars. When the up-type quarks and the down-type
quarks are allowed simultaneously to couple to more than one scalar
doublet, the diagonalization of the up-type and down-type mass
matrices does not automatically ensure the diagonalization of the
couplings with each single scalar doublet. Frequently, as in the Weinberg
model for CP violation \cite{Weinberg} or in Supersymmetry, the 2HDM
scalar potential and Yukawa lagrangian are  constrained by
a {\it ad hoc\/} discrete symmetry \cite{glash}, whose only role is
to protect the model from FCNC's at the tree level. Let us consider a
Yukawa lagrangian of the form

\ber
{\cal L}^{(A)}_{Y}= \eta^{U}_{ij} \bar Q_{i,L} \tilde\phi_1 U_{j,R} +
\eta^D_{ij} \bar Q_{i,L}\phi_1 D_{j,R} + 
\xi^{U}_{ij} \bar Q_{i,L}\tilde\phi_2 U_{j,R}
+\xi^D_{ij}\bar Q_{i,L} \phi_2 D_{j,R} \,+\, h.c. 
\eer

\noindent where $\phi_i$, for $i=1,2$, are the two scalar doublets of
a 2HDM, while $\eta^{U,D}_{ij}$ and $\xi_{ij}^{U,D}$ are the 
non-diagonal matrices of the Yukawa couplings.  

 When  no discrete symmetry
is imposed  then both up-type and down-type quarks can have FC
couplings \cite{PS}. Such models were called Class A in \cite{P}
 to be contrasted
with models with a forced absence of FCNC, called Class B.

In the notation and basis of Ref\cite{soni} the flavor changing
part of the lagrangian can be written as

\ber
{\cal L}_{Y,FC}^{(III)} = \hat\xi^{U}_{ij} \bar Q_{i,L}\tilde\phi_2 U_{j,R} 
 +\hat\xi^D_{ij}\bar Q_{i,L} \phi_2 D_{j,R} \,+\, h.c. \label{lyukfc}
\eer

\noindent where $Q_{i,L}$, $U_{j,R}$, and $D_{j,R}$ denote now the
quark mass eigenstates and $\hat\xi_{ij}^{U,D}$ are the rotated
couplings, in general not diagonal. If we define $V_{L,R}^{U,D}$ to be
the rotation matrices acting on the up- and down-type quarks, with
left or right chirality respectively, then the neutral FC couplings
will be

\ber
\hat\xi^{U,D}_{\rm neutral}=(V_L^{U,D})^{-1}\cdot \xi^{U,D}
\cdot V_R^{U,D} \,\,\,.
\eer

\noindent On the other hand for the charged FC couplings we will have

\ber
\hat\xi^{U}_{\rm charged}\!&=&\!\hat\xi^{U}_{\rm neutral}\cdot 
V_{{\rm CKM}}\nonumber\\
\hat\xi^{D}_{\rm charged}\!&=& \!V_{\rm {CKM}}\cdot
\hat\xi^{D}_{\rm neutral} \
\eer 

\noindent where $V_{{\rm CKM}}$ denotes the
Cabibbo-Kobayashi-Maskawa matrix. 

The phenomenology for the 2HDM, for the quarkonium processes under study,
 is not very different 
from the other two models considered above and so we will concentrate
mainly on the top color models and supersymmetry models with
R-parity violation.

\section{Results and Discussion of Theoretical Uncertainties}
\begin{table}[h]

\caption[]{\label{table1} \small Input parameters used in our calculations.}
\centering
\begin{tabular}{|c|c|}
\hline
\hline
NRQCD matrix elements &  Value \\
\hline
$\langle  O_1^\psi(^3S_1)  \rangle \approx 3
\langle \psi |O_1(^3S_1) |\psi \rangle$ &  0.73 GeV$^3$ \\
$\langle \Upsilon |O_1(^3S_1) |\Upsilon \rangle$ & 2.3 GeV$^3$ 
 \\
$\langle  O_8^\psi(^3S_1)  \rangle$ &  0.014 GeV$^{3}$
 \\
$\langle  O_8^\psi (^1S_0)  \rangle 
\approx \langle  O_8^\psi (^3P_0)  
\rangle /m_c^2$ & $10^{-2}\, {\rm GeV}^3$  \\
$\langle \Upsilon |O_8(^3S_1) |\Upsilon \rangle$
& $5 \times 10^{-4}$ GeV$^3$  \\
$\langle  O_8^\Upsilon (^1S_0)  \rangle 
\approx \langle  O_8^\Upsilon (^3P_0)  
\rangle /m_b^2$ & $7 \times 10^{-3}\, {\rm GeV}^3$  \\
\hline
\end{tabular}
\end{table}

 In this section we discuss the results of our calculations. First let 
us look at the $\Upsilon$ decays.
The inputs to our calculation are the various well known
 NRQCD matrix elements given in 
table.1 \cite{cheung,cho}. The standard model contribution 
to the branching ratio is
$5.2 \times 10^{-11}$ from the penguin induced $ b \to s$  transition.  The 
process $\Upsilon \to {\overline B} X_s$ can also have a contribution in the 
standard model from tree level processes. The effective Hamiltonian
, suppressing the Dirac structure of the currents, 
\ber
H_W & = & \frac{G_F}{\sqrt{2}}
        V_{ub}V_{us}^*\left[a_1({\overline u}b)({\overline s}u) +
a_2({\overline s}b)({\overline u}u)\right] \
\eer
where $a_1$ and $a_2$ are the QCD coefficients  can generate the process
$\Upsilon \to B^+ K^-$. 
We can estimate the branching ratio for this process as
$$ BR[\Upsilon \to B^+ K^-] \approx |\frac{V_{ub}}{V_{cb}}|^2
BR[\Upsilon \to B_c^+ K^-]$$
Using $ BR[\Upsilon \to B_c^+ K^-]$ calculated in Ref\cite{indianguys}
one obtains $BR[\Upsilon \to B^+ K^-] \sim 1.5 \times 10^{-14}$.
For a rough estimate of $BR[\Upsilon \to B^+ X_s]$ we can 
scale $BR[\Upsilon \to B^+ K^-]$ by the factor 
$BR[B \to D^0 X]/BR[B \to D \pi]$. 
The measured value of $BR[B \to D^0 X]$ 
\cite{PRD} includes $D^0$ coming from the decay of $D^{0*}$ and
$D^{+*}$. From the spin phase factors $BR[B \to D^* X] \sim 3 BR[B \to D X]$.
Hence $BR[B \to D^0 X]/BR[B \to D \pi] \sim 20$ leading to
$BR[\Upsilon \to B^+ X_s] \sim 3 \times 10^{-13}$.  
 
So far we have not considered $R_1$ and $R_2$ in $L_{new}$. 
 In our model independent analysis
we vary $R_1/\Lambda^2$, $R_2/\Lambda^2$ one at a time 
 and the use the 
constraint from measurements of $ b \to s e^+ e^{-}$ and
$ b \to s \mu^+ \mu^{-}$ . We 
 identify $\Lambda$ with the masses of the 
exchange particles which we take to be between $100-200$ GeV. We also take 
the cut-off for the integral in Eqs. 17-20 as $200$ GeV.
 The allowed values of
$R_1/\Lambda^2$, $R_2/\Lambda^2$  are then used to calculate
$\Upsilon \to {\overline B} X_s$
The constraint from $b \to s l^+ l^-$ gives
$$|R_{1,2}|/\Lambda^2 < (6-9)\times 10^{-6} (1/GeV)^2$$
Using the upper bounds on $|R_{1,2}|/\Lambda^2$ 
we find the branching ratio for the process  
$\Upsilon(1S) \to {\overline{B}} X_s$ 
to be between $(1-2) \times 10^{-6}$.  Branching ratios 
of similar order  are also obtained for
$\Upsilon(2S)$ and $\Upsilon(3S)$. For $\Upsilon(4S)$ the branching ratio is 
smaller by a factor of $100$ because of the larger width of
$\Upsilon(4S)$ which decays predominantly to two $B$ mesons.
 
Turning now to models, we find for the top color model from Eq. (24)
we can write
\ber
D^*_{Rbs} & = & 4\frac{R_1}{\Lambda^2}\frac{f_{\tpi}^2m_H^2}{m_t^2
|D_{Lbb}|^2D_{Rbb}}
\eer
We can identify $\Lambda=m_H$ and use the constraint from
$ b \to s e^+ e^{-}$ for a typical value of 
$|R_1|/\Lambda^2 \sim 6 \times 10^{-6} (1/GeV)^2$ .  
Assuming $|D_{Lbb}| \approx |D_{Rbb}| \approx 1$, and
$f_{\tpi}=50$GeV we obtain
$|D_{Rbs}| \sim 2 m_H^2 \times 10^{-6}$. With typical values of 
$m_H \sim 100-200$ GeV we get
$|D_{Rbs}| \sim 0.02-0.08$. Similar values have been obtained 
for $|D_{Rbs}|$ in Ref\cite{burdman} by considering the contributions of
 the charged higgs and top-pion  to $b \to s \gamma$. 
A similar exercise can be carried out with $|D_{Lbs}|$. Note
 that $B_s$ mixing probes
the combination $D^*_{Lbs}D_{Rbb}D^*_{Rbs}D_{Lbb}$ and so by either choosing
$R_1 \sim 0$ or $R_2 \sim 0$ we can satisfy the constraint on $B_s$mixing by 
choosing the appropriate mixing elements to be small. Note that in top color
models we can have operators 
${\overline s}(1-\gamma^5)b{\overline d} 
(1+\gamma^5) d $ and 
${\overline s}(1+\gamma^5)b{\overline d}
(1-\gamma^5) d $ that can contribute to
$ \Upsilon  \rightarrow {\overline B}   s {\overline d} \rightarrow
{\overline B} X_s $ 
after Fierz reordering. 
However these operators will be suppressed by form factor effects and also from mixing effects. We have checked that the contribution to
$ \Upsilon  \rightarrow {\overline B}  X_s $ from these operators are
much suppressed relative to the contribution of 
 the operators in $L_{new}$.  We will therefore not consider the the above operators in our analysis.

Turning to R-parity violating susy we first collect the 
constraints on the relevant couplings.
The upper limits of the $L$-violating couplings for the squark mass of
100 GeV are given by
\begin{eqnarray}
\vert \lambda^{\prime}_{kij}\vert&<&0.012,~(k,j=1,2,3; i=1,2),\\
\vert \lambda^{\prime}_{13j}\vert&<&0.16,~(j=1,2),\\
\vert \lambda^{\prime}_{133}\vert&<&0.001,\\
\vert \lambda^{\prime}_{23j}\vert&<&0.16,~(j=1,2,3),\\
\vert \lambda^{\prime}_{33j}\vert&<&0.26,~(j=1,2,3),
\end{eqnarray}
The first set of constraints in Eq. (38) come from the decay $K\rightarrow \pi \nu
\nu$ with FCNC processes in the down quark sector
\cite{agashe}. The 
set of constraints in Eq. (39) and Eq. (41) are obtained 
from the semileptonic decays of 
$B$-meson \cite{29}. 
The  constraint,
 on the coupling $\lambda^{\prime}_{133}$ in Eq. (40)  is obtained from the 
Majorana mass that the coupling can generate for the 
electron type neutrino \cite{21}.
The last set of limits in Eq. (42)  are derived from the leptonic decay modes of 
the $Z$ \cite{30}.
Assuming all the couplings to be positive we find the branching ratio for 
$\Upsilon \to {\overline B} X_s$ to be around $2 \times 10^{-6}$
for $m_{\tilde {\nu}}=100$GeV.

Turning next to $J/\psi \to {\overline D} X_u$, we first make an 
estimate for this process in the standard model. Since the penguin 
$ c \to u$ transition is
small in the standard model we neglect its contribution. As in the 
case for the
$\Upsilon$ system , for a rough estimate,  can write
$$ BR[ J/\psi \to D^0 X_u] 
\sim BR[ J/\psi \to D^0 \pi^0]
BR[D^0 \to K^- X]/BR[D^0 \to K^- \pi^+]$$  We obtain
$BR[ J/\psi \to D^0 \pi^0]$ from \cite{indianguys} and keeping in mind that
$BR[D^0 \to K^- X]$ contains contributions from states decaying to $K^-$
 we obtain  
 $BR[ J/\psi \to D^0 X_u] \sim 10^{-10}$ . A similar exercise gives
 $BR[ J/\psi \to D^+ X_u] \sim 10^{-9}$.

Considering new physics effects we can constrain $R_1$ and $R_2$
from $ c \to u l^+ l^-$. We get an estimate of the
 constraint on 
$c \to u e^+e^{-}$ 
by adding up the exclusive modes
$$ BR[D \to u e^+e^{-}] \ge BR[ D \to (\pi^0 + 
\eta + \rho^0 +\omega) e^+e^{-}]$$ 
From $c \to u l^+ l^-$ one obtains
$$|R_{1,2}|/\Lambda^2 \le 3.7 \times 10^{-4} (1/GeV)^2$$
 We find the branching fraction for the process 
$J/\psi \to {\overline D} X_u $
using the constraint from
$ c \to u l^+ l^-$  can be
$(3-4)\times 10^{-5}$

 In top color models  taking $R_1$ and $R_2$ one at a time, one obtains
$$\frac{2.1 \times 10^3}{{m_{\tpi}^4}}||U_{Lcc}|^2U_{Rtc}U^*_{Rtu}|^2 $$
 or
$$\frac{2.1 \times 10^3}{m_{\tpi}^4} ||U_{Rtc}|^2U_{Lcc}U^*_{Lcu}|^2 $$
as  
the branching fraction 
for $J/\psi \to {\overline D} X_u$. For $m_{\tpi}$ between
$100-200$ GeV this rate can be between $(0.1-2.0) \times 10^{-5}$
if all the mixing angles are $\sim 1$.
It has been shown in 
Ref\cite{kominis2} that our choices for $f_{\tpi}$ and $m_{\tpi}$
gives unacceptably large corrections to $Z \to b{\overline b}$ from
one loop
contribution of the top pions. However in a strongly coupled theory higher
loop terms can have significant contributions. Nonetheless 
if  we change
 $f_{\tpi}$ to $\sim 100$ GeV for better agreement with
$Z \to b {\overline b}$ data then the effect in
$J/\psi \to {\overline D} X_u$  is reduced by a factor of 16. As in the case of the $\Upsilon$ system we can satisfy the constraint from D mixing by choosing
$R_1 \sim 0$ or $R_2 \sim 0$.
 
For R parity violating susy, contribution to $J/\psi \to {\overline D} X_u$
 can only occur at loop level,
 with both the $\lambda^{\prime}$ or $\lambda^{\prime \prime}$ contributing,
through the box diagram
 and so is suppressed.
However in the general 2HDM, from Eq. 33 , the operator
${\overline u}c {\overline c}c$ can be generated by the tree level exchange
of the field $\phi_2$. The contribution to
$J/\psi \to {\overline D} X_u$ will be proportional to the
combination of couplings 
$ R_2 =\hat\xi^{U}_{12}\hat\xi^{U *}_{22}$ and
$R_1=\hat\xi^{U*}_{21}\hat\xi^{U }_{22}$.
 Note the $D^0-{\overline D^0}$ mixing 
probes $\hat\xi^{U}_{12}\hat\xi^{U*}_{21}$. So we can 
satisfy the constraint on $D^0-{\overline D^0}$ mixing  by choosing either
$R_1 \sim 0$ or 
$R_2 \sim 0$.
One then obtains
$$\frac{14}{m_{H}^4}|\hat\xi^{U}_{12}\hat\xi^{U *}_{22}|^2 $$
 or
$$\frac{14}{m_{H}^4}  |\hat\xi^{U*}_{21}\hat\xi^{U }_{22}|^2 $$
as  
the branching ratio for $J/\psi \to {\overline D} X_u$.
For $m_{H}$ between
$100-200$ GeV this rate can be between $(0.1-1.4) \times 10^{-7}$
if all the couplings are $\sim 1$. In a strongly interacting theory these
couplings can be $> 1$ as in the example of top color
models discussed above.

We note in passing that the four quark operators 
we have considered can also give rise to mixing operators 
that are $1/\Lambda^4$ suppressed. 
Taking one operator at a time one can generate the following operators that
contribute to mixing
\ber
O_1 & = & -\frac{3R_1^2}{2 \pi^2 \Lambda^2}\frac{m_b^2}{\Lambda^2}
\log \left(\frac{\Lambda^2}{m_b^2}\right)
{\overline {s}}(1-\gamma^5)b {\overline{s}}(1-\gamma^5)b, \nonumber\\
O_2 & = & -\frac{3R_2^2}{2 \pi^2 \Lambda^2}\frac{m_b^2}{\Lambda^2}
\log \left(\frac{\Lambda^2}{m_b^2}\right)
{\overline {s}}(1+\gamma^5)b {\overline{s}}(1+\gamma^5)b . \
\eer
We then have for $B_s$ mixing,
\ber
\Delta B_s & = & \frac{5}{3}{f_{B_s}}^2 \eta B M_{B_s}\delta , \
\eer
where $\eta$ is the QCD correction factor and $B$ and $\delta$
are defined through
$$
<B_s^0|{\overline {s}}(1-\gamma^5)b {\overline{s}}(1-\gamma^5)
|{\overline{B_s}^0}> = -\frac{5}{3}{f_{B_s}}^2  B {M_{B_s}}^2 ,\\
$$
and
$$
\delta  = -\frac{3R_i^2}{2 \pi^2 \Lambda^2}\frac{m_b^2}{\Lambda^2}
\log \left(\frac{\Lambda^2}{m_b^2}\right), \
$$
where $i=1,2$. A similar result can be written for $D$ mixing.

For the Upsilon system if we include the constraint from  $B_s$ mixing
generated by only the operators $O_{1,2}$  then
we obtain
$|R_{1,2}|/\Lambda^2 > 1-2 \times 10^{-6}(1/GeV)^2$. 
This is of the same order as the constraint obtained from $b \to s l^+ l^-$. 
In the case of $J/\psi$, if we include the constraint from the $D$ mixing
generated by only the operators $O_{1,2}$  then
we obtain
$|R_{1,2}|/\Lambda^2 \sim  10^{-6}(1/GeV)^2$. This will lower 
the branching fraction for
$J/\psi \to {\overline D} X_u $ to $\sim 10^{-9}$.
However,to evaluate consistently the effects of these operators at order of
${( 1 / \Lambda^2 )}^2$ 
would require the addition of other operators with
dimension $\le 8$.
%  in the effective Lagrangian that can contribute to mixing. 
As noted earlier we restrict our analysis to only dimension six operators 
in the effective lagrangian, thus the conservative result for
$J/\psi \to {\overline D} X_u $ is of order of $10^{-5}$, which, as we shown above, 
lies also in the region
predicted by the TopColor and 2HDM.
%so we have not included the
%mixing constraints
% from $O_{1,2}$ in our analysis.

Direct CP violation is possible in these decays through the interference
of the standard model and new physics contribution. The CP
 conserving phase is generated at the quark level from the penguin
diagrams. However the standard model contribution is small and so
the CP asymmetry $a_{CP}$ is also small with a typical value of $0.1 \%$
for $\Upsilon \to {\overline B} X_s$.

In summary we have calculated branching ratios for the flavor changing processes
$\Upsilon \to {\overline B} X_s$ and
$J/\psi \to {\overline D} X_u$. In a model independent description of
 new physics\cite{wang},
constrained by low energy data from $b \to s l^+ l^{-}$ and
$c \to u l^+l^{-}$, we found branching fractions for these processes can be 
$\sim 10^{-6}$ and $\sim 10^{-5}$ for $\Upsilon$ and $J/\psi$ decays 
respectively.
We also discussed several models of new physics that can 
allow these processes to
occur with branching ratios that maybe measurable in the next round of experiments.

\section{Acknowledgements}

We thank Amarjit Soni and Sean Fleming
 for useful discussions and comments. This work was
supported in part by the United States Department of Energy (S. Pakvasa),
NSF of China (X. Zhang) and by the Natural  Sciences and Engineering  Council
of Canada ( A. Datta and P. J. O'Donnell).

\end{document}